\documentclass[11 pt,twoside]{article}
\usepackage{amsmath,amsfonts,amssymb}
\usepackage{latexsym}
\usepackage{dcolumn}
\usepackage{graphicx,epsfig}
\usepackage{amsthm}
\usepackage{color}
\usepackage{framed}
\usepackage{cleveref}
\usepackage{cite}

\crefmultiformat{equation}{(#2#1#3)}%
       { and~(#2#1#3)}{, (#2#1#3)}{ and~(#2#1#3)}
\crefname{equation}{Eq.}{Eqs.}

\begin{document}

\setcounter{page}{1}

\pagestyle{plain} \vspace{0.5 cm}

\begin{center}
{\large{\bf {Quantum Corrections to the Accretion onto a Schwarzschild Black Hole in the Background of Quintessence}}}\\

\vspace{1 cm}

{\bf Kourosh Nozari$^{\dag,}$}\footnote{knozari@umz.ac.ir},\quad {\bf Milad Hajebrahimi$^{\dag,}$}\footnote{m.hajebrahimi@stu.umz.ac.ir}, \quad and \quad
{\bf Sara Saghafi$^{\dag,}$}\footnote{s.saghafi@stu.umz.ac.ir}\\

\vspace{0.5 cm}

$^{\dag}$Department of Theoretical Physics, Faculty of Basic Sciences, University of Mazandaran,\\
P. O. Box 47416-95447, Babolsar, Iran\\

\end{center}

\vspace{0.5 cm}

\begin{abstract}
It is well known that quantum effects may lead to remove the intrinsic singularity point of back holes. Also, the quintessence scalar field is a candidate model for describing late-time acceleration expansion. Accordingly, Kazakov and Solodukhin considered the existence of back-reaction of the spacetime due to the quantum fluctuations of the background metric to deform Schwarzschild black hole, which led to change the intrinsic singularity of the black hole to a 2-sphere with a radius of the order of the Planck length. Also, Kiselev rewrote the Schwarzschild metric by taking into account the quintessence field in the background. In this study, we consider the quantum-corrected Schwarzschild black hole inspired by Kazakov-Solodukhin's work, and Schwarzschild black hole surrounded by quintessence deduced by Kiselev to study the mutual effects of quantum fluctuations and quintessence on the accretion onto the black hole. Consequently, the radial component of 4-velocity and the proper energy density of the accreting fluid have a finite value on the surface of its central 2-sphere due to the presence of quantum corrections. Also, by comparing the accretion parameters in different kinds of black holes, we infer that the presence of a point-like electric charge in the spacetime is somewhat similar to some quantum fluctuations in the background metric.\\
\\
\textbf{PACS:} 97.10.Gz, 04.20.Dw, 04.60.-m, 04.60.Bc, 04.70.-s, 04.70.Dy, 95.36.+x\\
\textbf{Key Words:} Accretion Disk, Singularities, Black Hole Physics, Quintessence.
\end{abstract}

\newpage

\enlargethispage{\baselineskip}
\tableofcontents

\section{Introduction}\label{intro}

The existence of essential singularities (which leads to various black hole spacetimes) is one of the major problems in general relativity and it seems to be a common property in most of the solutions of Einstein’s field equations. It is commonly believed that a successful quantization of gravity will provide us with modifications to the theory which are necessary to avoid the prediction of geodesically incomplete spacetime manifolds \cite{Duff1974,Frolov1981,Modesto2004,Modesto2011,Gambini2008,Nicolini2005}. Quantum corrections may completely change the gravitational equations and the corresponding spacetime geometry at the Planck scale. Kazakov and Solodukhin in 1994 took into account the influence of quantum corrections on the behaviour of the Schwarzschild solution. This solution is probably the most important one in general relativity, when quantum corrections are taken into account. They added the outcomes of quantum field theories into the Einstein-Hilbert action within the general theory of relativity, which led to remove the point-like singularity of the Schwarzschild (SCH) black hole. They took into account the back-reaction of the spacetime due to the quantum fluctuations of the background metric (see more details in \cite{Kazakov1994}). Consequently, they found the quantum-corrected Schwarzschild (KS) black hole in which there is a central 2-dimensional sphere with a radius of the order of the Planck length, instead of a central point-like singularity. Therefore, despite the original SCH black hole, the central region in this modified black hole has a volume, which is equal to the Planck volume. Recently, the studying of the Hawking radiation via tunneling process for this modified black hole is investigated in Ref. \cite{Hajebrahimi2020}. Also, the spectroscopy of the KS black hole is studied in Ref. \cite{Shahjalal2019a}. Here, we consider accretion of matter onto this type of black hole. We will discuss the critical fluid velocity, the speed of sound, the mass accretion rate, and so on.

On the other hand, a number of astronomical observations confirmed that the late-time universe is undergoing an accelerated expansion, such as observations of type Ia supernova \cite{Perlmutter1999,Riess1998,Garnavich1998,Riess1999,Lapuente1995,Riess2004}. In order to explain this observed phenomenon, an unknown energy component, dubbed dark energy, must have been introduced in the framework of general relativity, looking only on the matter part of Einstein's equations \cite{Scranton2003,Sherwin2011,Cabr2006}. The simplest candidate for the dark energy is the cosmological constant, which is consistent with most of the current astronomical observations; however, it suffers from the cosmological constant problem \cite{Weinberg1989,Carroll2001,Bull2016} and maybe the age problem \cite{Yang2010} as well. It is thus natural to consider other complicated cases. A dynamic scalar field can also serve as the dark energy component such as quintessence, phantom, $\kappa$-essence, etc. Quintessence is the simplest scalar field dark energy model without having theoretical problems like Laplacian instabilities or ghosts \cite{Carroll1998,Zlatev1999,Sahni2000,Matos2001,Capozziello2006}. One Schwarzschild-like solution related to the quintessence model was found in \cite{Kiselev2003} by Kiselev. The solution describes a spherically symmetric and static exterior spacetime surrounded by a quintessence field. To derive this solution, Kiselev used the quintessence stress-energy tensor with the additivity and linearity conditions as $T_{tt}=T_{rr}=\rho_{q}$ and $T_{\theta \theta}=T_{\phi \phi}=-\big(\rho_{q}\big(3\omega_{q}+1\big)/2\big)$. There are some notable papers in the literature focused on the Kiselev (KSL) black hole, such as the study of its null and time-like geodesics structure \cite{Fernando2012,Uniyal2015}, the study of geodesics structure of Schwarzschild-anti de Sitter black hole with quintessence \cite{Malakolkalami2015}, the study of accretion onto Reissner-Nordstr\"{o}m-anti-de Sitter black hole surrounded by quintessence \cite{Zheng2019}, the study of gravitational lensing due to this Schwarzschild-like black hole \cite{Younas2015} and etc. \cite{Ghaderi2016,Malakolkalami2015,Pedraza2020,Jiao2017}. In 2019, Shahjalal took into account the KS and KSL black holes, simultaneously and found the metric of the quantum-corrected Schwarzschild black hole surrounded by the quintessence (we call it as Kazakov-Solodukhin-Kiselev (KSK) black hole), and then its thermodynamics properties, in addition to its area and entropy quantization was investigated \cite{Shahjalal2019b,Shahjalal2019ab}. Recently, the geodesic structure and Hawking radiation as tunneling process for such a black hole is studied in Refs. \cite{Nozari2020,Eslamzadeh2020}. Also, effects of quantum corrections on the criticality and efficiency of such black holes are studied in Ref. \cite{Bezerra2019}. In Ref. \cite{Jiao2017}, the authors considered accretion onto a SCH black hole with static exterior spacetime surrounded by a quintessence field (i.e., onto the KSL black hole).

Accretion of matter onto black hole is an important phenomenon of long-standing interest to astrophysicists. Pressure-free gas being dragged onto a massive central object was examined in \cite{Hoyle1939,Bondi1944}. This study was generalized to the case of the spherically symmetric and transonic accretion of adiabatic fluids onto astrophysical objects \cite{Bondi1952}. The steady-state spherical symmetric flow of matter into or out of a condensed object in the framework of general relativity was investigated in \cite{Michel1972}. Issues of the critical points of accretion were discussed in \cite{Begelman1978,Malec1999}. Accretion has been analyzed in various cases in the literature, including accretion onto the SCH black hole, onto a Kerr-Newman black hole \cite{Babichev2008,Madrid2008,Bhadra2012}, onto a Reissner-Nordstr\"{o}m (RN) black hole, onto the Reissner-Nordstr\"{o}m-de Sitter (RNdS) black hole, onto the KSL black hole \cite{Jiao2017} and etc. \cite{Ganguly2014,Mach2013a,Mach2013b,Karkowski2013,Salahshoor2018}.

In this paper, we found it interesting to check the accretion onto the KSK black hole in which we can explore how the quantum fluctuations of the background metric and the dark energy ingredient of the Universe affect and modify the previously known outcomes of the general theory of relativity. The rest of the paper is organized as follows. In the next section, we investigate the accretion onto the KS black hole in which we first introduce this quantum-corrected black hole, and then we derive the basic equations and dynamical parameters for accretion onto that. Also we will determine the critical points and the conditions they must fulfill. In Section \ref{AoKSKBH}, we study the accretion onto the KSK black hole in which we determine the equations and dynamical parameters for accretion onto the hole and as in Section \ref{AoKSBH}, the critical points and conditions are evaluated. Section \ref{CaC} contains some comparison between the achieved outcomes in Section \ref{AoKSBH} and Section \ref{AoKSKBH} with the results arisen from accretion onto the SCH black hole, onto the RN black hole, onto the RNdS black hole, and onto the KSL black hole. In this section, the main focus would be on the quantum fluctuations of the background metric along with the dark energy ingredient of the Universe and how they can affect and modify the accretion equations and parameters. Finally, we will briefly summarize and discuss our results in Section \ref{SaC}.

\section{Accretion onto the KS Black Hole}\label{AoKSBH}

\subsection{Introduction to KS Black Hole}\label{ItKSBH}

Following what performed in Ref. \cite{Kazakov1994}, it is possible to change the central point-like singularity of the SCH black hole to a central 2-dimensional spherical region with radius $a$ which is of the order of the Planck length $r\sim l_{Pl}\equiv a$ to gain the KS black hole. Through mathematical computations (see in Ref. \cite{Kazakov1994}), the line element of the KS black hole can be found as follows
\begin{equation}\label{dsfr}
ds^{2}=f\left(r\right)dt^{2}-\frac{dr^{2}}{f\left(r\right)}-r^2\left(d\theta^{2}+\sin^{2}{\theta}\,d\phi^{2}\right)\,,
\end{equation}
where
\begin{equation}\label{fr}
f\left(r\right)=-\frac{2M}{r}+\frac{1}{r}\sqrt{r^{2}-a^{2}}\,,
\end{equation}
and $M$ is the black hole mass (in the unit of $G=c^{2}=1$). Due to the small value of the quantum correction parameter, $a$, one can expand Eq. \eqref{fr} for large $r$ (i.e., $r\gg a$) to find the following relation \cite{Hajebrahimi2020}
\begin{equation}\label{frexpand}
f(r)\approx 1-\frac{2M}{r}-\frac{a^{2}}{2r^{2}}\,,
\end{equation}
remembering the RN black hole.

In order to find the horizons of the KS black hole, it is required to compute the roots of $f(r)$ in Eq. \eqref{frexpand}, i.e., $f(r)=0\,,$ which are as follows
\begin{equation}\label{routin}
r_{\pm}=\frac{1}{2}\left[2M\pm\sqrt{4M^{2}+2a^{2}}\right]\,.
\end{equation}
Heading from the outside of the event horizon to the region between two horizons, the behaviour of the $t$-coordinate ($r$-coordinate), will change from time-like (space-like) to space-like (time-like). Then, crossing from the region between two horizons to the central 2-sphere with radius $a$, the behaviour of the $t$-coordinate ($r$-coordinate) will change from space-like (time-like) to time-like (space-like). Therefore, in the KS black hole, $r_{-}$ is the inner/Cauchy horizon, and $r_{+}$ is the event horizon, completely similar to the RN black hole.

\subsection{Basic Dynamical Equations}\label{BDEks}

Now, we introduce the basic equations of the accretion onto the KS black hole. We consider the perfect fluid inflow matter onto the black hole. The stress-energy tensor of the perfect fluid matter in the framework of the general relativity is as follows
\begin{equation}\label{tmunu}
T^{\mu\nu}=\left(p+\rho\right)u^{\mu}u^{\nu}-pg^{\mu\nu}\,,
\end{equation}
where $\rho$ is the proper energy density, $p$ is the proper pressure, and $u^{\mu}$ is the 4-velocity of the fluid, and also $g^{\mu\nu}$ is the metric tensor. To flow the fluid radially in the equatorial plane, we suppose that $\theta=\frac{\pi}{2}$. Hence, the 4-velocity of the fluid only has two component associated with $t$ and $r$ coordinates.
Using the normalization condition of the 4-velocity, i.e. $u_{\mu}u^{\mu}=1$, one can find the $t$-component of the 4-velocity of the fluid as follows
\begin{equation}\label{ut}
u^{\left(t\right)}=\frac{\sqrt{1-\frac{2M}{r}-\frac{a^{2}}{2r^{2}}+\left(u^{\left(r\right)}\right)^{2}}}
{\left(1-\frac{2M}{r}-\frac{a^{2}}{2r^{2}}\right)}\,.
\end{equation}
For an inward flow i.e., the accretion, it is necessary that $u^{(r)}<0$.

The conservation of the stress-energy tensor of the perfect fluid results in the following relation
\begin{equation}\label{a0}
\left(p+\rho\right)u^{\left(r\right)}r^2\sqrt{1-\frac{2M}{r}-\frac{a^{2}}{2r^{2}}+\left(u^{(r)}\right)^{2}}=A_{0}\,,
\end{equation}
where the integration constant is defined as $A_{0}$. One can project the conservation of the stress-energy tensor onto the 4-velocity to acquire the following equation
\begin{equation}\label{umutmunumu}
u_{\mu}T^{\mu\nu}_{;\nu}=0\,.
\end{equation}
After some straightforward computations, Eq. \eqref{umutmunumu} yields
\begin{equation}\label{rhod}
\frac{\rho'}{\left(p+\rho\right)}+\frac{\left(u^{\left(r\right)}\right)'}{u^{\left(r\right)}}+\frac{2}{r}=0\,,
\end{equation}
where ($'$) means the differentiation with respect to $r$. From Eq. \eqref{rhod}, one can find
\begin{equation}\label{a1}
r^{2}u^{\left(r\right)}\exp\left[\int\frac{d\rho}{p+\rho}\right]=-A_{1}\,,
\end{equation}
where the integration constant is defined as $A_{1}$. As mentioned above, we know that $u^{\left(r\right)}<0$. Therefore, one can write
\begin{equation}\label{a2}
\left(p+\rho\right)\sqrt{1-\frac{2M}{r}-\frac{a^{2}}{2r^{2}}+\left(u^{\left(r\right)}\right)^{2}}\,\exp\left[-\int\frac{d\rho}{p+\rho}\right]=A_{2}\,,
\end{equation}
where $A_{2}$ is an integration constant.

In this framework, the mass flux equation can be written as follows \cite{Salahshoor2018}
\begin{equation}\label{rhoumumu}
\left(\rho u^{\mu}\right)_{;\mu}=0\,,
\end{equation}
which leads to the following result
\begin{equation}\label{a3}
\rho u^{\left(r\right)}r^{2}=A_{3}\,,
\end{equation}
where $A_{3}$ is a constant of integration.

\subsection{Dynamical Parameters}\label{DPks}

In this setup, we choose the isothermal fluids which have a constant temperature. Such fluids can flow at a constant temperature \cite{Salahshoor2018}. Hence, their equation of state is of the form $p=\omega\rho$ in which $\omega$ is the equation of state parameter. So, Eqs. \cref{a1,a2,a3} lead to the following result
\begin{equation}\label{a4}
\frac{\left(p+\rho\right)}{\rho}\sqrt{1-\frac{2M}{r}-\frac{a^{2}}{2r^{2}}+\left(u^{\left(r\right)}\right)^{2}}=A_{4}\,,
\end{equation}
where $A_{4}$ is a constant of integration. One can substitute $p=\omega\rho$ in Eq. \eqref{a4} to find $u^{(r)}$ as follows
\begin{equation}\label{urr}
u^{\left(r\right)}=\left(\frac{1}{\omega+1}\right)\sqrt{A_{4}^{2}-\left(\omega+1\right)^{2}
\left(1-\frac{2M}{r}-\frac{a^{2}}{2r^{2}}\right)}\,.
\end{equation}
Also, from Eq. \eqref{a3} the proper energy density can be deduced as follows
\begin{equation}\label{rhoa3}
\rho=\left(\frac{A_{3}}{r^{2}}\right)\frac{\left(\omega+1\right)}{\sqrt{A_{4}^{2}-\left(\omega+1\right)^{2}
\left(1-\frac{2M}{r}-\frac{a^{2}}{2r^{2}}\right)}}\,.
\end{equation}

\subsection{Mass Evolution}\label{MEks}

In the framework of black hole physics, it is well known that the mass of the black hole cannot be fixed, but it changes over time. Accretion disks around black holes cause the black hole mass to increase by accreting mass onto it. On the other hand, the Hawking radiation makes the black hole to lose mass.

One can integrate the flux of the fluid over the surface of the KS black hole, $\mathcal{S}$ to obtain its mass rate through the accretion process (for more details see Ref. \cite{Salahshoor2018}) in the form  $\dot{M}\equiv\frac{dM}{dt}=-\int T^{r}_{t}d\mathcal{S}$, where $d\mathcal{S}=\big(\sqrt{-g}\big)d\theta d\phi$ is the black hole surface element and $g$ is the determinant of the metric tensor, $g_{\mu\nu}$. From Eq. \eqref{tmunu}, we have $T^{r}_{t}=\left(p+\rho\right)u_{t}u^{r}$. Consequently, we obtain the mass rate as
\begin{equation}\label{mda0}
\dot{M}=-4\pi r^{2}u^{\left(r\right)}\left(p+\rho\right)\sqrt{1-\frac{2M}{r}-\frac{a^{2}}{2r^{2}}+\left(u^{\left(r\right)}\right)^{2}}\equiv -4\pi A_{0}\,,
\end{equation}
where $A_{0}=-A_{1}A_{2}$ and $A_{2}=\left(p_{\infty}+\rho_{\infty}\right)\sqrt{f\left(r_{\infty}\right)}$. So, one can write \cite{Salahshoor2018}
\begin{equation}\label{mdinfty}
\dot{M}=4\pi A_{1}M^{2}\left(p_{\infty}+\rho_{\infty}\right)\sqrt{f\left(r_{\infty}\right)}\,.
\end{equation}
From Eq. \eqref{mdinfty}, one can deduce a relation for the mass of the black hole as a function of its initial mass, $M_{i}$ and $t$ as follows \cite{Salahshoor2018}
\begin{equation}\label{mtmi}
M\left(t\right)=\frac{M_{i}}{1-\frac{t}{t_{cr}}}\,,
\end{equation}
where the critical time is defined as $t_{cr}\equiv\left[4\pi A_{1}M_{i}\left(p+\rho\right)\sqrt{f\left(r_{\infty}\right)}\right]^{-1}$. In the case $t=t_{cr}$ the denominator of
\cref{mtmi} vanishes and the black hole mass grows up to infinity in a finite time.

\subsection{Critical Accretion}\label{CAks}

One can differentiate Eqs. \cref{a3,a4} with respect to $r$ to find the following two relations
\begin{equation}\label{rrdur1}
\frac{\rho'}{\rho}+\frac{\left(u^{\left(r\right)}\right)'}{u^{\left(r\right)}}+\frac{2}{r}=0\,,
\end{equation}
\begin{equation}\label{rdrlnd2}
\frac{\rho'}{\rho}\bigg(\frac{d\ln[p+\rho]}{d\ln[\rho]}-1\bigg)+\frac{u^{(r)}\big(u^{(r)}\big)'}
{\left(1-\frac{2M}{r}-\frac{a^{2}}{2r^{2}}+\big(u^{(r)}\big)^{2}\right)}+\frac{2Mr+a^{2}}{2r^{3}\left(1-\frac{2M}{r}-\frac{a^{2}}{2r^{2}}+\left(u^{\left(r\right)}\right)^{2}\right)}=0\,.
\end{equation}
Eqs. \eqref{rrdur1} and \eqref{rdrlnd2} yield
\begin{equation}\label{dlnur}
\frac{d\ln\left[u^{\left(r\right)}\right]}{d\ln[r]}=\frac{\alpha_{1}}{\alpha_{2}}\,,
\end{equation}
where we have defined $\alpha_{1}$ and $\alpha_{2}$ as follows
\begin{equation}\label{alpha1}
\alpha_{1}\equiv-2V^{2}+\frac{2Mr+a^2}{2r^{2}\left(1-\frac{2M}{r}-\frac{a^{2}}{2r^{2}}+\left(u^{\left(r\right)}\right)^{2}
\right)}\,,
\end{equation}
\begin{equation}\label{alpha2}
\alpha_{2}\equiv V^{2}-\frac{\left(u^{\left(r\right)}\right)^{2}}{\left(1-\frac{2M}{r}-\frac{a^{2}}{2r^{2}}+\left(u^{\left(r\right)}\right)^{2}
\right)}\,.
\end{equation}
From Eqs. \cref{alpha1,alpha2}, one can find the following result
\begin{equation}\label{v2}
V^{2}\equiv\frac{d\ln\left[p+\rho\right]}{d\ln\left[\rho\right]}-1\,.
\end{equation}

To find critical points, it is necessary to satisfy $\alpha_{1}=\alpha_{2}=0$ as a condition. Such a condition leads to the following relations
\begin{equation}\label{vcr}
V_{cr}^{2}=\frac{2Mr+a^{2}}{r^{2}\left[4\left(1-\frac{2M}{r}-\frac{a^{2}}{2r^{2}}\right)+\frac{2Mr+a^{2}}{r^{2}}\right]}\,,
\end{equation}
\begin{equation}\label{urcr}
u^{(r)}_{cr}=\frac{2Mr+a^{2}}{4r^{2}}\,.
\end{equation}
Since the left-hand side of Eq. \eqref{vcr} is always positive, the right-hand side must be positive, too. Hence, the range of critical radius can be obtained by solving the following inequality
\begin{equation}\label{cr1}
4\left(1-\frac{2M}{r}-\frac{a^{2}}{2r^{2}}\right)+\frac{2Mr+a^{2}}{r^{2}}>0\,.
\end{equation}
From Eqs. \cref{a4,urr}, one can compute the sound speed, $c_{s}^{2}=\frac{dp}{d\rho}$ as follows
\begin{equation}\label{csa4}
c_{s}^{2}=-1+A_{4}\sqrt{1-\frac{2M}{r}-\frac{a^{2}}{2r^{2}}+\left(u^{\left(r\right)}\right)^{2}}\,.
\end{equation}
The accreting matter flow needs to cross over the critical radius, $r=r_{cr}$ as tends to the black hole. So, the radial element of the 4-velocity of the fluid matches the local sound speed, i.e., $u_{cr}=c_{s}$.

\section{Accretion onto the KSK Black Hole}\label{AoKSKBH}

\subsection{Introduction to the KSK Black Hole}\label{ItKSKBH}

As what performed in Refs. \cite{Shahjalal2019b, Nozari2020}, it is possible to write the line element of the KSK black hole by combining the KS black hole with the KSL one to gain the following line element
\begin{equation}\label{dsgr}
ds^{2}=g\left(r\right)dt^{2}-\frac{dr^{2}}{g\left(r\right)}-r^{2}\left(d\theta^{2}+\sin^{2}{\theta}\,d\phi^{2}\right)\,,
\end{equation}
where
\begin{equation}\label{gr1}
g\left(r\right)=-\frac{2M}{r}+\frac{1}{r}\sqrt{r^{2}-a^{2}}-\frac{\sigma}{r^{3\omega_{q}+1}}\,,
\end{equation}
in which $M$ is the black hole mass (in the unit $G=c^{2}=1$), and $\sigma$ is a positive normalization factor associated with the quintessence field \cite{Kiselev2003}. Again we can expand $g(r)$ in Eq. \eqref{gr1} for large $r$ to find
\begin{equation}\label{gr1expand}
g\left(r\right)\approx 1-\frac{2M}{r}-\frac{a^{2}}{2r^{2}}-\frac{\sigma}{r^{3\omega_{q}+1}}\,.
\end{equation}

As previously mentioned, the quintessence's parameter of the equation of state is in the range $-1\leq\omega_{q}\leq-\frac{1}{3}$. The case $\omega_{q}=-1$, which is known as the cosmological constant model, results in the following relation for $g\left(r\right)$ in Eq. \eqref{gr1expand}
\begin{equation}\label{gr3rnksk}
g\left(r\right)=1-\frac{2M}{r}-\frac{a^{2}}{2r^{2}}-\sigma r^{2}\,.
\end{equation}
We call the line element corresponding to Eq. \eqref{gr3rnksk} as Reissner-Nordstr\"{o}m-Kazakov-Solodukhin-Kiselev (RNKSK) black hole. Obviously, this is similar to the metric coefficient of the RNdS black hole. The properties and a review on the Cauchy horizon, entropy, gravitational lensing effects, the study of the Hawking radiation as tunneling, strong cosmic censorship, and the thermodynamics of the RNdS black hole has been investigated, respectively in Refs. \cite{Kowalczyk2006, Chambers1997, Zhang2016, Ren2002, Zhao2016, Jiang2006, Mo2018, Ge2019, Liu2019, Zhang2011, Manna2020}.

Another case is choosing $\omega_{q}=-\frac{2}{3}$ in Eq. \eqref{gr1expand}, which leads to
\begin{equation}\label{gr2expand}
g\left(r\right)=1-\frac{2M}{r}-\frac{a^{2}}{2r^{2}}-\sigma r\,,
\end{equation}
and we call the black hole corresponding to the metric coefficient in Eq. \eqref{gr2expand} as Ideal-Kazakov-Solodukhin-Kiselev (IKSK) black hole. One can find the roots of $g\left(r\right)$ in Eq. \eqref{gr2expand} to obtain the horizons of the IKSK black hole as follows
\begin{equation}\label{rin}
r_{in}=-\frac{Y}{6\sqrt[3]{2}\,\sigma}+\frac{1}{3\sigma}+X\,,
\end{equation}
\begin{equation}\label{reve}
r_{eh}=\frac{\left(1-\mathrm{i}\sqrt{3}\right)Y}{12\sqrt[3]{2}\,\sigma}+\frac{1}{3\sigma}-\frac{1}{2}\left(1+\mathrm{i}
\sqrt{3}\right)X\,,
\end{equation}
\begin{equation}\label{rcosmo}
r_{ch}=\frac{\left(1+\mathrm{i}\sqrt{3}\right)Y}{12\sqrt[3]{2}\,\sigma}+\frac{1}{3\sigma}-\frac{1}{2}\left(1-\mathrm{i}
\sqrt{3}\right)X\,,
\end{equation}
where ``eh'' stands for ``event horizon'', ``ch'' stands for ``cosmological horizon'', and we have defined
\begin{equation}\label{Ydef}
Y=\left(\sqrt{\left(108a^{2}\sigma^{2}+144\sigma M-16\right)^{2}+4(24\sigma M-4)^{3}}+108a^{2}\sigma^{2}+144\sigma M-16\right)^{\frac{1}{3}}\,,
\end{equation}
and
\begin{equation}\label{Xdef}
X=\frac{24\sigma M-4}{3\sqrt[3]{4}\,\sigma Y}\,.
\end{equation}
By checking the properties of the coordinates around these horizons, one can deduce that $r_{in}$ is an inner/Cauchy horizon, like the RN black hole's inner/Cauchy horizon, $r_{eh}$ is an event horizon, and $r_{ch}$ is a cosmological or the quintessence horizon, like the cosmological horizon of de Sitter spacetime. Such a horizon also exists in the KSL black hole with $\omega_{q}=-\frac{2}{3}$. Between the event horizon and the cosmological one, the spacetime is static. In what follows, we use the KSK metric coefficient in Eq. \eqref{gr1expand} to derive the equations and physical parameters of the accretion. Then, in Section \ref{CaC}, we investigate and check the results by setting $\omega_{q}=-1, -\frac{2}{3}$ (corresponding to RNKSK and IKSK cases, respectively) just for examples, while we know that they have not a special advantage than other values in the range $-1\leq\omega_{q}\leq-\frac{1}{3}$.

\subsection{Basic Dynamical Equations}\label{BDEksk}

 Now we suppose the perfect fluid inflow of matter onto the KSK black hole. Following the procedure of the previous section, one can find the time-component of the 4-velocity of the perfect fluid as
\begin{equation}\label{utut}
u^{\left(t\right)}=\frac{\sqrt{1-\frac{2M}{r}-\frac{a^{2}}{2r^{2}}-\frac{\sigma}{r^{3\omega_{q}+1}}
+\left(u^{\left(r\right)}\right)^{2}}}{1-\frac{2M}{r}-\frac{a^{2}}{2r^{2}}-\frac{\sigma}{r^{3\omega_{q}+1}}}\,.
\end{equation}
In addition, accretion means an inward flow of fluid onto the black hole, which needs negative values for radial-component of the 4-velocity of the fluid, $u^{\left(r\right)}$.

From the conservation of the stress-energy tensor, one can find the following relation
\begin{equation}\label{b0}
\left(p+\rho\right)u^{\left(r\right)}r^2\sqrt{1-\frac{2M}{r}-\frac{a^{2}}{2r^{2}}-\frac{\sigma}{r^{3\omega_{q}+1}}
+\left(u^{(r)}\right)^{2}}=B_{0}\,,
\end{equation}
where $B_{0}$ is an integration constant. By projecting the conservation of the stress-energy tensor onto the 4-velocity in Eq. \eqref{umutmunumu}, one can write
\begin{equation}\label{b1}
r^{2}u^{\left(r\right)}\exp\left[\int\frac{d\rho}{p+\rho}\right]=-B_{1}\,,
\end{equation}
where $B_{1}$ is an integration constant. As mentioned above, we know that $u^{\left(r\right)}<0$, thus we can rewrite Eq. \eqref{b0} as follows
\begin{equation}\label{b2}
\left(p+\rho\right)\sqrt{1-\frac{2M}{r}-\frac{a^{2}}{2r^{2}}-\frac{\sigma}{r^{3\omega_{q}+1}}
+\left(u^{\left(r\right)}\right)^{2}}\,\exp\left[-\int\frac{d\rho}{p+\rho}\right]=B_{2}\,,
\end{equation}
where $B_{2}$ is an integration constant. Also, from the mass flux relation in Eq. \eqref{rhoumumu}, one can find
\begin{equation}\label{b3}
\rho u^{\left(r\right)}r^{2}=B_{3}\,,
\end{equation}
where $B_{3}$ is an integration constant.

\subsection{Dynamical Parameters}\label{DPksk}

For isothermal fluids with the equation of state of the form $p=\omega\rho$ ($\omega$ is the parameter of this equation of state), and based on Eq. \eqref{a4}, one can write the following relation
\begin{equation}\label{b4}
\frac{\left(p+\rho\right)}{\rho}\sqrt{1-\frac{2M}{r}-\frac{a^{2}}{2r^{2}}-\frac{\sigma}{r^{3\omega_{q}+1}}
+\left(u^{\left(r\right)}\right)^{2}}=B_{4}\,,
\end{equation}
where $B_{4}$ is a integration constant. By substituting $p=\omega\rho$ into Eq. \eqref{b4}, one can find $u^{\left(r\right)}$ as follows
\begin{equation}\label{uromeg}
u^{\left(r\right)}=\left(\frac{1}{\omega+1}\right)\sqrt{B_{4}^{2}-\left(\omega+1\right)^{2}\left(1-\frac{2M}{r}-
\frac{a^{2}}{2r^{2}}-\frac{\sigma}{r^{3\omega_{q}+1}}\right)}\,,
\end{equation}
and consequently, the proper energy density of the fluid to the form
\begin{equation}\label{rhob3b4}
\rho=\left(\frac{B_{3}}{r^{2}}\right)\frac{\left(\omega+1\right)}{\sqrt{B_{4}^{2}-\left(\omega+1\right)^{2}
\left(1-\frac{2M}{r}-\frac{a^{2}}{2r^{2}}-\frac{\sigma}{r^{3\omega_{q}+1}}\right)}}\,.
\end{equation}

As the previous section, one can write the relevant equations for the mass evolution through the accretion onto the KSK black hole correspond to \cref{mda0,mdinfty,mtmi} just by substituting $B_{j}$ for $A_{j}$ where $(j=0,1,2, 3)$, and $g(r)$ for $f(r)$ in the corresponding equations.

\subsection{Critical Accretion}\label{CAksk}

Based on what performed in the previous section, one can deduce
\begin{equation}\label{bet1bet2}
\frac{d\ln\left[u^{\left(r\right)}\right]}{d\ln[r]}=\frac{\beta_{1}}{\beta_{2}}\,,
\end{equation}
where
\begin{equation}\label{bet1}
\beta_{1}\equiv -2W^{2}+\frac{2Mr+a^{2}+\sigma\left(3\omega_{q}+1\right)r^{1-3\omega_{q}}}{2r^{2}\left(1-\frac{2M}{r}-\frac{a^{2}}{2r^{2}}
-\frac{\sigma}{r^{3\omega_{q}+1}}+\left(u^{\left(r\right)}\right)^{2}\right)}\,,
\end{equation}
\begin{equation}\label{bet2}
\beta_{2}\equiv W^{2}-\frac{\left(u^{\left(r\right)}\right)^{2}}{\left(1-\frac{2M}{r}-\frac{a^{2}}{2r^{2}}-\frac{\sigma}{r^{3\omega_{q}+1}}
+\left(u^{\left(r\right)}\right)^{2}\right)}\,,
\end{equation}
and
\begin{equation}\label{w2}
W^{2}\equiv\frac{d\ln\left[p+\rho\right]}{d\ln\left[\rho\right]}-1\,.
\end{equation}
Now, critical points occur when $\beta_{1}=\beta_{2}=0$, which leads to the following relations
\begin{equation}\label{wcr2}
W_{cr}^{2}=\frac{2Mr+a^{2}+\sigma\left(3\omega_{q}+1\right)r^{1-3\omega_{q}}}{r^{2}\left(4\left(1-\frac{2M}{r}-
\frac{a^{2}}{2r^{2}}-\frac{\sigma}{r^{3\omega_{q}+1}}\right)+\frac{2Mr+a^{2}+\sigma\left(3\omega_{q}+1\right)r^{1-3\omega_{q}}}
{r^{2}}\right)}\,,
\end{equation}
and
\begin{equation}\label{urcr2}
u^{\left(r\right)}_{cr}=\frac{2Mr+a^{2}+\sigma\left(3\omega_{q}+1\right)r^{1-3\omega_{q}}}{4r^{2}}\,.
\end{equation}
Also, the range of the critical radius is shown by the following condition
\begin{equation}\label{crcrcr}
4\left(1-\frac{2M}{r}-\frac{a^{2}}{2r^{2}}-\frac{\sigma}{r^{3\omega_{q}+1}}\right)+\frac{2Mr+a^{2}+\sigma\left(3\omega_{q}+
1\right)r^{1-3\omega_{q}}}{r^2}>0\,.
\end{equation}
Finally, the sound speed, $c_{s}^{2}=\frac{dp}{d\rho}$ is found as
\begin{equation}\label{csb4}
c_{s}^{2}=-1+B_{4}\sqrt{1-\frac{2M}{r}-\frac{a^{2}}{2r^{2}}-\frac{\sigma}{r^{3\omega_{q}+1}}
+\left(u^{\left(r\right)}\right)^{2}}\,.
\end{equation}
Now, we can consequently find the trace of the dark energy ingredient of the Universe as well as quantum correction effects in the sound speed expression.

\section{Comparison and Consequences}\label{CaC}

In this section, we first compare the results of radial element of the 4-velocity and the proper energy density of the accreting perfect fluid in KS black hole, which is studied in Section \ref{AoKSBH} with their corresponding outcomes of RN and SCH black holes. Then, we compare the results of these two physical quantities in RNKSK black hole with $\omega_{q}=-1$ and IKSK black hole with $\omega_{q}=-\frac{2}{3}$ with their corresponding outcomes of KSL (from Ref. \cite{Jiao2017}), RNdS, and SCH black holes.

\subsection{Radial-Component of 4-Velocity}\label{RC4V}

Arbitrarily, one can set $\omega=\frac{1}{2}$ as the parameter of the equation of state of the isothermal fluid in Eq. \eqref{urr} to deduce the radial-component of 4-velocity of the accreting fluid for KS black hole as
\begin{equation}\label{ura}
u^{(r)}=\left(\frac{2}{3}\right)\sqrt{A_{4}^{2}-\frac{9}{4}\left(1-\frac{2M}{r}-\frac{a^{2}}{2r^{2}}\right)}\,.
\end{equation}
Replacing the metric coefficient of KS black hole with RN one, which is $\tilde{f}(r)=\left(1-\frac{2M}{r}+\frac{Q^{2}}{r^{2}}\right)$ in the above relation, results in the radial-component of 4-velocity of the accreting fluid onto the RN black hole as $$u^{(r)}=\left(\frac{2}{3}\right)\sqrt{\tilde{A}_{4}^{2}-\frac{9}{4}\left(1-\frac{2M}{r}+\frac{Q^{2}}{r^{2}}\right)}\,,$$
in which $\tilde{A}_{4}$ is an integration constant in RN case correspond to $A_{4}$, and $Q$ is the electric charge of the RN black hole. Obviously, we see that the radial-component of 4-velocity in KS black hole is completely equivalent to the corresponding relation in RN black hole, by setting $$Q\equiv\mathrm{i}\frac{\sqrt{2}}{2}a\,.$$ This outcome encourages us to infer that the electric charge acts similar to the quantum effects in this study, as the result in Ref. \cite{Hajebrahimi2020}. Such a result is interesting in the literature, this is also shown in Ref. \cite{Nozari2006} in which one can deduce that a commutative RN black hole is equivalent to a non-commutative Schwarzschild black hole, thermodynamically. Therefore, it seems that the presence of a point-like electric charge in the spacetime results in some quantum fluctuations in the background metric. So, by considering this outcome, one can probe the quantum gravitational impacts with the electrical effects, and also, find a more meaningful concept for a point-like electric charge in the spacetime.

In addition, for SCH black hole, one can write the radial-component of 4-velocity of the accreting fluid as follows $$u^{(r)}=\left(\frac{2}{3}\right)\sqrt{K_{4}^{2}-\frac{9}{4}\left(1-\frac{2M}{r}\right)}\,,$$ where $K_{4}$ is the corresponding integration constant for SCH black hole. The radial-component of 4-velocity in RN and SCH black holes at $r=0$ becomes infinite, i.e., the accreting matter falls into point-like singularity and this physical quantity reaches an infinite value, which is not sensible or acceptable. But, it is possible to compute the radial-element of the 4-velocity on the surface of the central 2-sphere in KS black hole by setting $r=a$ in Eq. \eqref{ura} which is a finite value. Hence, it seems that in this setup, the accreting matter can cross the event horizon to accrete onto central 2-sphere without falling into a point-like singularity. Such an interesting result is completely due to the presence of quantum effects in this setup, which leads to remove the central singularity of the SCH black hole, so that it would change the inner structure of the black hole.

\begin{figure}
\centering
\includegraphics[width=0.7\textwidth]{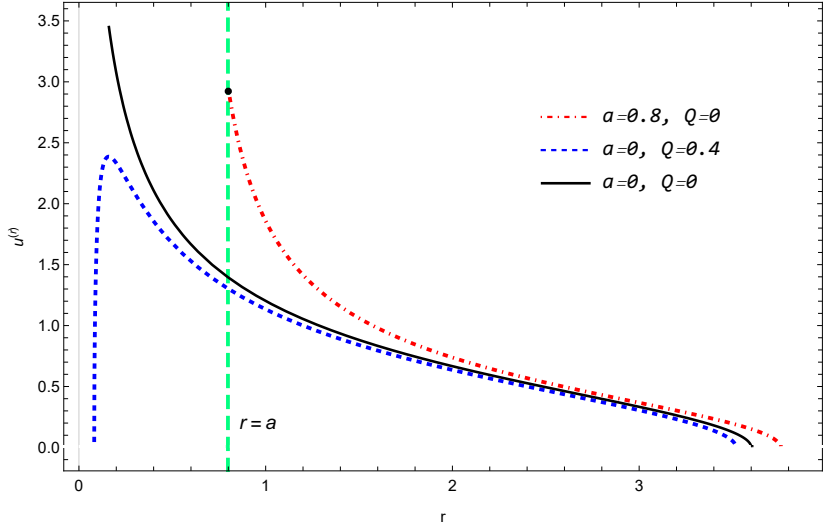}
\caption{\label{Fig1}\small{\emph{The illustration of the function $u^{(r)}$ versus $r$ for KS, RN, and SCH black holes in which we have set $M=1$, $A_{4}=\tilde{A}_{4}=K_{4}=1$, $Q=0.4$, and $a=0.8$. The red dot-dashed line, the thick black line, and the blue dashed line are for KS, RN, and SCH cases, respectively. Also, the green dashed line shows the surface of the central $2$-sphere in this two dimensional figure. The choice of $a=0.8$ is just for having a better illustration, while we know that $a$ is of the order of the Planck length.}}}
\end{figure}

Figure \ref{Fig1} shows the behavior of the radial-component of 4-velocity of the accreting fluid for KS, RN, and SCH black holes. From this figure, we see that this physical quantity on the surface of the central 2-sphere of the KS black hole has a finite physical value, and also is bounded in this value. But in the SCH black hole, the radial-element goes to an infinite value by approaching the point-like singularity of this black hole at $r=0$. In the case of RN black hole, the radial-component reaches a maximum and then falls to zero, but then suddenly tends to infinite value by approaching the singularity of the charged black hole. So, one can deduce that the presence of quantum effects in KS case leads to regularize the inner structure of the black hole to prevent the falling of accreting matter into a point-like singularity.

Now, supposing $\omega=\frac{1}{2}$ for the parameter of the equation of state of the isothermal perfect fluid in Eq. \eqref{uromeg}, one can find the radial-component of 4-velocity of the accreting fluid for KSK black hole as
\begin{equation}\label{urkskrnksk}
u^{\left(r\right)}=\left(\frac{2}{3}\right)\sqrt{B_{4}^{2}-\frac{9}{4}\left(1-\frac{2M}{r}-
\frac{a^{2}}{2r^{2}}-\frac{\sigma}{r^{3\omega_{q}+1}}\right)}\,,
\end{equation}
and in what follows, we investigate two cases associated with $\omega_{q}=-1, -\frac{2}{3}$ corresponding to RNKSK and IKSK, respectively.

\subsubsection{RNKSK case with $\omega_{q}=-1$}\label{omegami1}

Inserting $\omega_{q}=-1$ in Eq. \eqref{urkskrnksk} results in
\begin{equation}\label{urrnksk}
u^{\left(r\right)}=\left(\frac{2}{3}\right)\sqrt{B_{4}^{2}-\frac{9}{4}\left(1-\frac{2M}{r}-
\frac{a^{2}}{2r^{2}}-\sigma r^{2}\right)}\,,
\end{equation}
which is the radial-element of 4-velocity for RNKSK black hole. Also, replacing the metric coefficient of the RNKSK black hole with the RNdS one, which is $\tilde{g}(r)=\left(1-\frac{2M}{r}+
\frac{Q^{2}}{r^{2}}-\frac{\Lambda}{3}r^{2}\right)$ results in the radial-component of 4-velocity for RNdS black hole as $$u^{\left(r\right)}=\left(\frac{2}{3}\right)\sqrt{\tilde{B}_{4}^{2}-\frac{9}{4}\left(1-\frac{2M}{r}+
\frac{Q^{2}}{r^{2}}-\frac{\Lambda}{3}r^{2}\right)}\,,$$ where $Q$ is the charge of the RNdS black hole, $\Lambda$ is the positive cosmological constant, and $\tilde{B}_{4}$ is the corresponding integration constant for RNdS black hole. The radial-component of the 4-velocity of the RNKSK and the RNdS black holes are equivalent, if we set $$Q\equiv\mathrm{i}\frac{\sqrt{2}}{2}a$$ and $$\Lambda\equiv 3\sigma\,.$$ This analogy means that the RNKSK black hole approximately behaves the same as the RNdS black hole. This is an expected result as we previously mentioned that the case $\omega_{q}=-1$, is known as the cosmological constant model. In this setup we see that the quintessence parameter, $\sigma$ is approximately equal to the cosmological constant, $\Lambda$. Again, it is clear that the electric charge in this setup mimics the quantum effects in the background spacetime. On the other hand, the radial-element of the 4-velocity in RNdS black hole and even KSL one with $\omega_{q}=-1$ (known from Ref. \cite{Jiao2017}) at $r=0$, where is the position of their central singularity, becomes infinite. But, this physical quantity in RNKSK black hole at $r=a$, i.e., on the surface of its central 2-sphere, has a finite value, similar to  the KS black hole.

\subsubsection{IKSK case with $\omega_{q}=-\frac{2}{3}$}\label{omegami2to3}

Setting $\omega_{q}=-\frac{2}{3}$ in Eq. \eqref{urkskrnksk} results in the radial-element of the 4-velocity for IKSK black hole as
\begin{equation}\label{urksk1}
u^{\left(r\right)}=\left(\frac{2}{3}\right)\sqrt{B_{4}^{2}-\frac{9}{4}\left(1-\frac{2M}{r}-
\frac{a^{2}}{2r^{2}}-\sigma r\right)}\,.
\end{equation}
The above relation again has a finite value on the surface of the central 2-sphere of this modified black hole. But, the radial element of 4-velocity of the accreting fluid in KSL black hole with $\omega_{q}=-\frac{2}{3}$, which is derived in Ref. \cite{Jiao2017}, has a infinite value at the location of the singularity of this black hole. Therefore, in KS, RNKSK, and IKSK cases, we see that the presence of quantum effects leads to have a finite physical value for the radial-component of the 4-velocity, which means that in these cases, the accreting matter circulates around the central 2-sphere, rather than falling into a point-like singularity.

\begin{figure}
\centering
\includegraphics[width=0.7\textwidth]{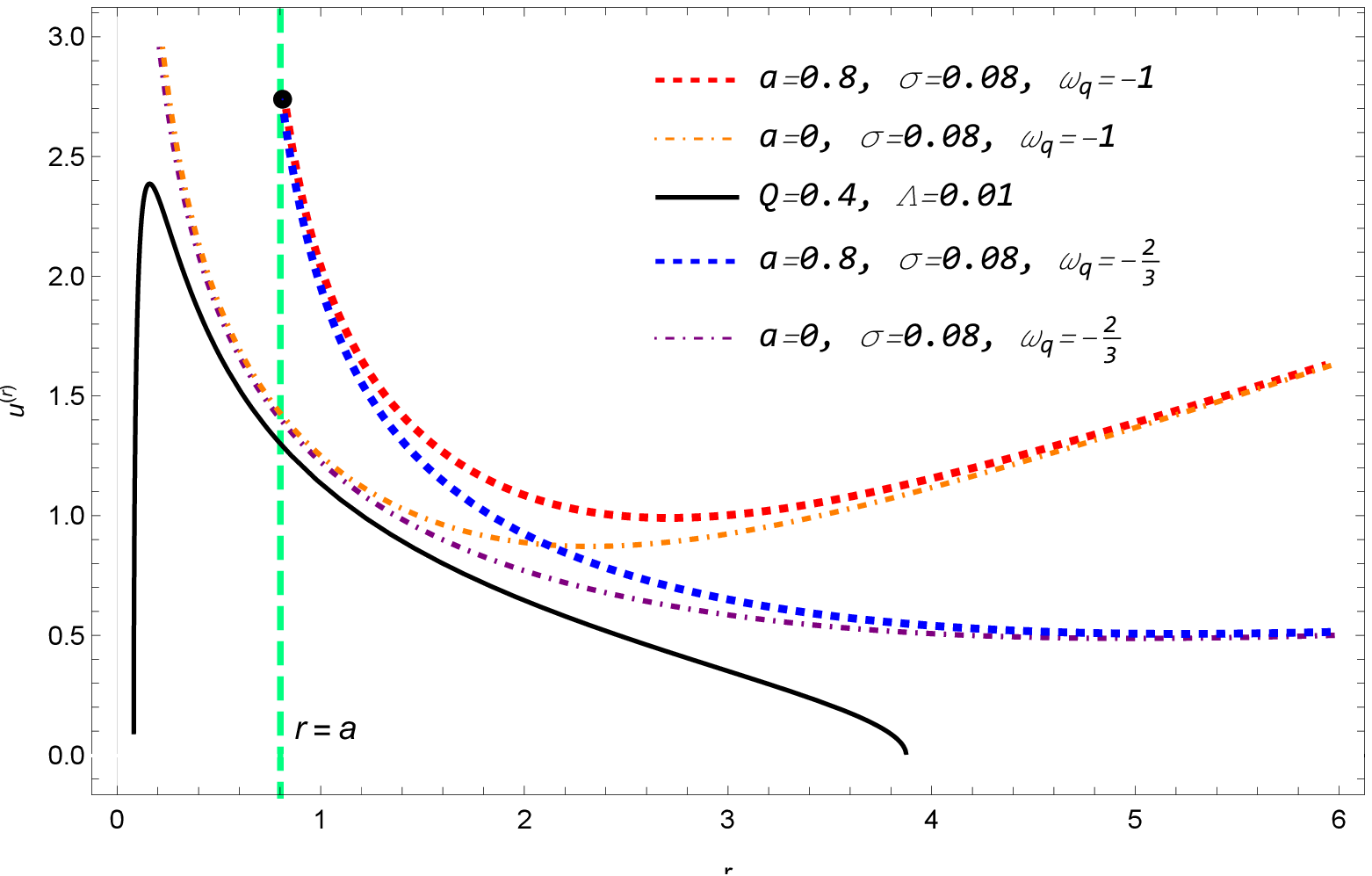}
\caption{\label{Fig2}\small{\emph{The illustration of the function $u^{\left(r\right)}$ versus $r$ for RNKSK, RNdS, IKSK, KSL with $\omega_{q}=-1$, and KSL with $\omega_{q}=-\frac{2}{3}$ cases in which we set $M=1$, $B_{4}=\tilde{B}_{4}=1$, $Q=0.4$, $\Lambda=0.01$, $\sigma=0.08$, and $a=0.8$. The red and blue dashed lines are for RNKSK and IKSK cases, respectively. The orange and purple dot-dashed lines are for KSL with $\omega_{q}=-1$, and KSL with $\omega_{q}=-\frac{2}{3}$ cases, respectively. The thick black line is the line of RNdS case. Finally, the green dashed line is the location of the surface of the central 2-sphere. The choice of $a=0.8$ is just for having a better illustration, while we know that $a$ is of the order of the Planck length.}}}
\end{figure}

Figure \ref{Fig2} depicts the curves of the radial-component of 4-velocity of the fluid in RNKSK, RNdS, IKSK, KSL with $\omega_{q}=-1$, and KSL with $\omega_{q}=-\frac{2}{3}$ black holes. This figure shows that the radial velocity of RNKSK and KSK black holes have been bounded on the surface of their central 2-sphere, and have a finite value on this surface. Also, this quantity in KSL with $\omega_{q}=-1$ and KSL with $\omega_{q}=-\frac{2}{3}$ cases, which can be found in Ref. \cite{Jiao2017} goes to an infinite value by approaching their singularities due to the absence of quantum effects. Figures \ref{Fig1} and \ref{Fig2} illustrate that the RNdS black hole behaves as the RN one. Also, Far from the black hole, it is clear that the radial-component of the RNKSK case approaches to the KSL case with $\omega_{q}=-1$, and the IKSK case approaches to the KSL with $\omega_{q}=-\frac{2}{3}$, respectively. This happens due to the intense quantum effects near the horizons and singularities of these black holes, far from black holes, these effects become negligible.

\subsection{Proper Energy Density}\label{PED}

Supposing $\omega=\frac{1}{2}$ as the parameter of the equation of state of the isothermal fluid in Eq. \eqref{rhoa3}, we derive the proper energy density of the accreting fluid for KS black hole as
\begin{equation}\label{rhoaks}
\rho=\frac{3A_{3}}{2r^{2}\sqrt{A_{4}^{2}-\frac{9}{4}\left(1-\frac{2M}{r}-\frac{a^{2}}{2r^{2}}\right)}}\,.
\end{equation}
Substituting the metric coefficient of RN black hole with KS one in the above relation results in the proper energy density of accreting fluid onto the RN black hole as $$\rho=\frac{3\tilde{A}_{3}}{2r^{2}\sqrt{\tilde{A}_{4}^{2}-\frac{9}{4}\left(1-\frac{2M}{r}+\frac{Q^{2}}{r^{2}}\right)}}\,,$$ in which $\tilde{A}_{3}$ and $\tilde{A}_{4}$ are integration constants in RN case corresponding to $A_{3}$ and $A_{4}$, respectively, and $Q$ is the electric charge of the RN black hole. Once again, the equivalence of the proper energy density in the KS and the RN black hole is obvious by letting $Q\equiv\mathrm{i}\frac{\sqrt{2}}{2}a$.

On the other hand, one can write the proper energy density of the fluid for the SCH black hole as $$\rho=\frac{3K_{3}}{2r^{2}\sqrt{K_{4}^{2}-\frac{9}{4}\left(1-\frac{2M}{r}\right)}}\,,$$ where $K_{3}$ and $K_{4}$ are the corresponding integration constants in SCH case. In both of the RN and the SCH black holes, the proper energy density becomes infinite at their singularity positions, i.e., $r=0$. But, the proper energy density of the accreting fluid in KS black hole at the surface of its central 2-sphere, i.e., $r=a$ has a finite physical value. This means that the accreting matter can cross its event horizon to approach the 2-sphere, and then circulate around it, rather than falling into a singularity. As in  Eq. \eqref{rhoaks}, the proper energy density is proportional to inverse of $r$, at first glance. So, in the KS case, at $r=a$ the proper energy density has a very large value due to the fact that the quantum correction parameter, $a$, is of the order of the Planck length \cite{Kazakov1994}.

\begin{figure}
\centering
\includegraphics[width=0.7\textwidth]{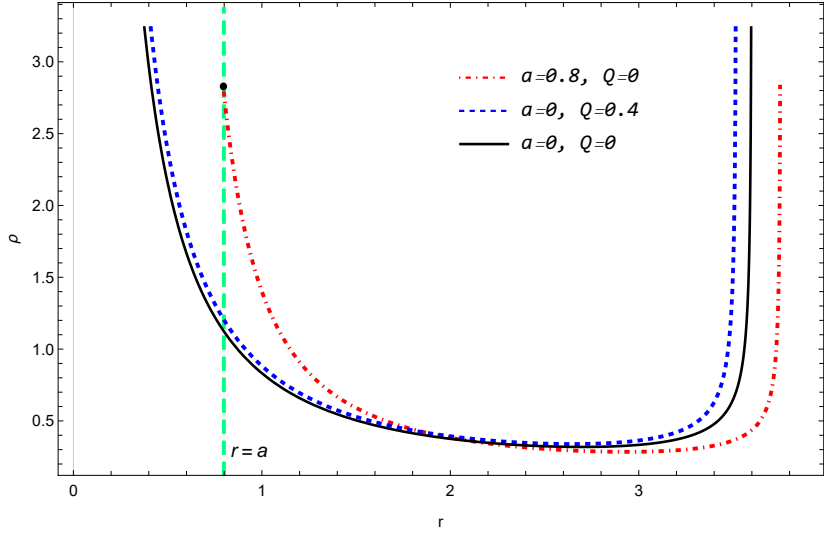}
\caption{\label{Fig3}\small{\emph{The illustration of the density function $\rho$ versus $r$ for KS, RN, and SCH black holes in which we have set $M=1$, $A_{3}=\tilde{A}_{3}=K_{3}=1$, $A_{4}=\tilde{A}_{4}=K_{4}=1$, $Q=0.4$, and $a=0.8$. The red dot-dashed line, the blue dashed line, and the thick black line are for KS, RN, and SCH case, respectively. Also, the green dashed line shows the surface of the central 2-sphere in this two dimensional figure at which the curve of the proper energy density in the KS black hole is bounded. The choice of $a=0.8$ is just for having a better illustration, while actually $a$ is of the order of the Planck length.}}}
\end{figure}

Figure \ref{Fig3} shows the proper energy density of the accreting fluid in KS, RN, and SCH black holes. We see from this figure that the proper energy density in the KS black hole is bounded at $r=a$, which is on the surface of the 2-sphere at the center of the black hole. Also, in this case, this physical quantity has a finite value, in spite of RN and SCH black holes in which the proper energy density goes to infinite, unlimitedly. So, quantum effects make the black hole regular, so that the accreting matter inside the event horizon do not fall into a singularity to reach an infinite proper energy density.

Now, again we suppose that $\omega=\frac{1}{2}$ to find the proper energy density of the accreting fluid in the KSK black hole as
\begin{equation}\label{rhoakskrnksk}
\rho=\frac{3B_{3}}{2r^{2}\sqrt{B_{4}^{2}-\frac{9}{4}\left(1-\frac{2M}{r}-\frac{a^{2}}{2r^{2}}-\frac{\sigma}{r^{3\omega_{q}+1}}
\right)}}\,.
\end{equation}
What follows is the investigation of RNKSK and IKSK cases associated with $\omega_{q}=-1, -\frac{2}{3}$ respectively.

\subsubsection{RNKSK case with $\omega_{q}=-1$}\label{omegami1gr}

Inserting $\omega_{q}=-1$ in Eq. \eqref{rhoakskrnksk} lets to find the proper energy density of accreting fluid in RNKSK black hole as $$\rho=\frac{3B_{3}}{2r^{2}\sqrt{B_{4}^{2}-\frac{9}{4}\left(1-\frac{2M}{r}-\frac{a^{2}}{2r^{2}}-\sigma r^{2}\right)}}\,.$$ One can again obtain the proper energy density in RNdS black hole as $$\rho=\frac{3\tilde{B}_{3}}{2r^{2}\sqrt{\tilde{B}_{4}^{2}-\frac{9}{4}\left(1-\frac{2M}{r}+\frac{Q^{2}}{r^{2}}-\frac{\Lambda}{3} r^{2}\right)}}\,,$$ where $Q$ is the charge of the RNdS black hole, $\Lambda$ is the positive cosmological constant, and also, $\tilde{B}_{3}$ and $\tilde{B}_{4}$ are the corresponding integration constat for RNdS black hole. Again by setting $Q\equiv\mathrm{i}\frac{\sqrt{2}}{2}a$ and $\Lambda\equiv 3\sigma$, one can see the equivalence of the RNdS black hole with the RNKSK one. At $r=0$, which is the location of the singularity of the RNdS black hole and KSL one with $\omega_{q}=-1$ (known from Ref. \cite{Jiao2017}), the proper energy density becomes infinite. But, in RNKSK black hole, we can see that at $r=a$ this physical quantity has a finite, but large value due to the smallness of the quantum correction parameter, $a$.

\subsubsection{IKSK case with $\omega_{q}=-\frac{2}{3}$}\label{omegami2to3gr}

Supposing $\omega_{q}=-1$ in Eq. \eqref{rhoakskrnksk} results in the proper energy density of accreting fluid in IKSK black hole as
\begin{equation}\label{rhoakskgr}
\rho=\frac{3B_{3}}{2r^{2}\sqrt{B_{4}^{2}-\frac{9}{4}\left(1-\frac{2M}{r}-\frac{a^{2}}{2r^{2}}-\sigma r\right)}}\,.
\end{equation}
The proper energy density of the accreting fluid in KSL black hole with $\omega_{q}=-\frac{2}{3}$, which can be seen in Ref. \cite{Jiao2017} has an infinite value at the position of the singularity of this black hole. But, one can conclude that the proper energy density in IKSK black hole has a finite value on the surface of its central 2-sphere of this black hole, i.e., $r=a$. Therefore, the presence of the quantum fluctuations in the KS, the IKSK, and the RNKSK black holes leads to produce a finite, physical large value for the proper energy density, which means that the accreting matter spreads and moves around the central 2-sphere, rather falling into a point-like singularity.

\begin{figure}
\centering
\includegraphics[width=0.7\textwidth]{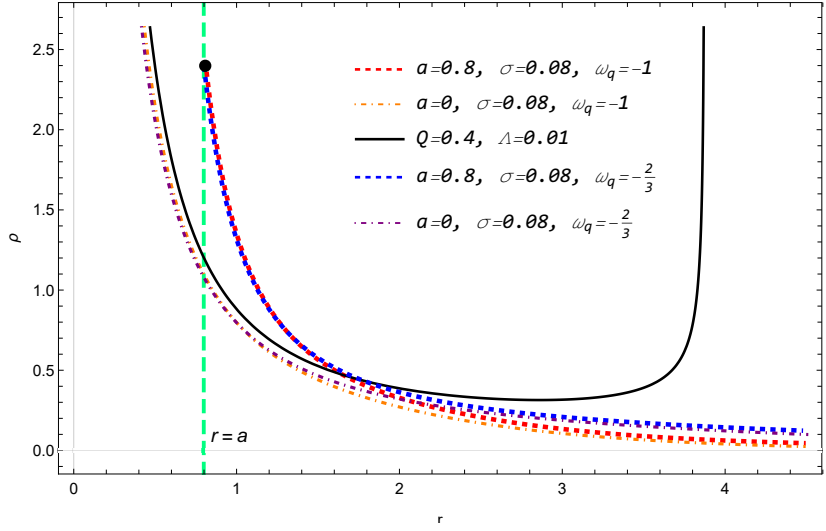}
\caption{\label{Fig4}\small{\emph{The illustration of $\rho$ versus $r$ for RNKSK, RNdS, IKSK, KSL with $\omega_{q}=-1$, and KSL with $\omega_{q}=-\frac{2}{3}$ black holes in which we have set $M=1$, $B_{3}=\tilde{B}_{3}=1$, $B_{4}=\tilde{B}_{4}=1$, $Q=0.4$, $\Lambda=0.01$, $\sigma=0.08$, and $a=0.8$. The red and blue dashed lines are for RNKSK and IKSK cases, respectively. The orange and purple dot-dashed lines are for KSL with $\omega_{q}=-1$, and KSL with $\omega_{q}=-\frac{2}{3}$ cases, respectively. The thick black line is for RNdS case. Finally, the green dashed line is the location of the surface of the central 2-sphere at which the RNKSK and IKSK cases is bounded. The choice of $a=0.8$ is just for having a better illustration, while actually $a$ is of the order of the Planck length.}}}
\end{figure}

Figure \ref{Fig4} depicts the curve of the proper energy density of the accreting fluid in RNKSK, RNdS, IKSK, KSL with $\omega_{q}=-1$, and KSL with $\omega_{q}=-\frac{2}{3}$ black holes. In cases RNKSK and IKSK, the proper energy density has a finite physical value, and also is bounded on the surface of the central 2-sphere, in spite of RNdS, KSL with $\omega_{q}=-1$, and KSL with $\omega_{q}=-\frac{2}{3}$ cases where this quantity goes to infinity by approaching their singularities. Also, in all of the illustrated cases, except RNdS case, increasing $r$ leads to the proper energy density approaching zero. So, the presence of quantum effects makes the inner structure of the black hole regular, and the presence of quintessence results in approaching the proper energy density zero at far from the black hole. Also, the proper energy density in RNdS black hole acts the same as RN and SCH cases.

\section{Summary and Conclusions}\label{SaC}

In this paper, we have considered accretion onto the quantum-corrected Schwarzschild black hole (KS black hole), and also, this quantum-corrected black hole surrounded by the quintessence field (KSK black hole). We obtained the fundamental equations for the flow of matter onto these types of black holes. We derived the general analytic expressions for the critical points and the mass accretion rate and determined the physical conditions the critical points should fulfill. Our main focus was on how the quantum fluctuations of the background metric can modify the accretion parameters and equations. For better understanding this point, some comparison was performed with other types of black holes. We have analyzed the behavior of the radial velocity and the proper energy density for different cases. We came to conclusion that on the contrary to the SCH, RN, KSL, and RNdS cases, in which the radial component of the 4-velocity and the proper energy density of the accreting fluid become infinite at the singularity ($r=0$) of these black holes, by considering the quantum corrections as fluctuations of the background metric in the KS and KSK (i.e. both RNKSK and IKSK) black holes, the mentioned parameters have a finite value at $r=a$, i.e., on the surface of its central 2-sphere. In other words, due to the presence of quantum effects in the KS and KSK (i.e. both RNKSK and IKSK) cases, which lead to consider a central 2-dimensional sphere with a radius of the order of the Planck length instead of a central point-like singularity in SCH black hole, the accretion process and parameters undergo dramatic changes toward regularization. Also, by considering the radial component of velocity of the accreting fluid in RN and KS black holes, we infer that the electric charge acts similar to the quantum effects in this investigation. Observation of interesting feature that the presence of a point-like electric charge in the spacetime results in some quantum fluctuations in the background metric in our context, resembles some previously reported works in this issue \cite{Hajebrahimi2020,Nozari2006}.\\
\\
{\bf Acknowledgement}\\
We would like to appreciate the respected referee for carefully reading the manuscript, for insightful comments and advise that have boosted the quality of the paper considerably.

\end{document}